\documentclass[reprint,floatfix,aps,prd,showpacs,footinbib,amsmath,amssymb,amsfonts,superscriptaddress]{revtex4-1}
\usepackage{bm}
\usepackage{graphicx}
\usepackage{color}
\usepackage{dsfont}
\usepackage[normalem]{ulem}
\usepackage{hyperref}
\usepackage{mathrsfs}
\usepackage{calrsfs}
\usepackage[utf8]{inputenc}
\usepackage{varioref}
\usepackage[active]{srcltx}
\newcommand{\BigO}[1]{\ensuremath{\operatorname{O}\bigl(#1\bigr)}}
\expandafter\ifx\csname package@font\endcsname\relax\else
\expandafter\expandafter
\expandafter\usepackage
\expandafter\expandafter
\expandafter{\csname package@font\endcsname}%
\fi
\hypersetup{
	colorlinks=true,       
	linkcolor=blue,          
	citecolor=blue,        
}

\usepackage[section]{placeins}

\usepackage{fancybox}
\labelformat{figure}{Fig.~#1} 

\begin{document}

\title{Quasinormal modes as a distinguisher between general relativity and $ f(R) $ gravity}


\author{Soham Bhattacharyya} 
\affiliation{School of Physics, Indian Institute of Science Education and Research Thiruvananthapuram (IISER-TVM), Trivandrum 695016, India}
\email{xeonese13@iisertvm.ac.in}
\author{S. Shankaranarayanan}
\affiliation{School of Physics, Indian Institute of Science Education and Research Thiruvananthapuram (IISER-TVM), Trivandrum 695016, India}
\affiliation{Department of Physics, Indian Institute of Technology-Bombay, Mumbai 400076, India}
\email{shanki@phy.iitb.ac.in}



\begin{abstract}
Quasi-Normal Modes (QNM) or ringdown phase of gravitational waves provide critical information about the structure of compact objects like Black Holes. Thus, QNMs can be a tool to test General Relativity (GR) and possible deviations from it. In the case of GR, it is known for a long time that a relation between two types of Black Hole perturbations: scalar (Zerilli) and vector (Regge-Wheeler), leads to an equal share of emitted gravitational energy. With the direct detection of Gravitational waves, it is now natural to ask: whether the same relation (between scalar and vector perturbations) holds for modified gravity theories? If not, whether one can use this as a way to probe deviations from General Relativity. As a first step, we show explicitly that the above relation between Regge-Wheeler and Zerilli breaks down for a general $ f(R) $ model and hence the two perturbations do not share equal amounts of emitted gravitational energy. We discuss the implication of this imbalance on observations and the no-hair conjecture.
\end{abstract}

\pacs{}

\maketitle

\section{Introduction}
General Relativity has been highly successful and has passed all the weak field and indirect strong field tests~\cite{1993-Will-Book,*2014-Will-LRR}. The existence of black-holes (BHs) is one of the key predictions of general relativity. It is argued that close to the BH singularity since the curvature is infinite, general relativity does not hold. In general, it has been suggested that at large curvatures, one need to include higher order curvature terms to the standard Einstein-Hilbert action~\cite{1977-Stelle-PRD,*1978-Stelle-GRG,Clifton:2011jh}. There are several candidates for such higher derivative additions like the contracted Riemann/Ricci Curvature or higher powers of Ricci scalar. The inclusion of such terms produces stabilization of divergence structure of gravity~\cite{1977-Stelle-PRD}. 

When a star collapses to a BH or two BHs merge to form another 
BH, the event horizon of the remnant BH is highly distorted and radiates gravitational waves until it settles down to an equilibrium configuration~\cite{1957-Regge.Wheeler-PR,*1970-Zerilli-PRL,*1962-Newman.Penrose-JMP,*Vishveshwara:1970zz,Chandrasekhar1984}. The gravitational radiation emitted is a superposition of damped sinusoidal; the frequency and the damping of these quasi-normal modes (QNMs) depend only on the parameters characterizing the BH (like Mass, Charge and angular momentum) and is independent of the initial configuration that caused the excitation~\cite{Nollert:1999ji,*1999-Kokkotas.Schmidt-LRR,*2011-Konoplya.Zhidenko-RMP}.
Hence, QNMs play a key role, as their detection would confirm the nature of the remnant BH. The first detection of the gravitational waves from the event GW150914 confirmed the emitted gravitational wave from the remnant BH characteristic of the QNM frequencies~\cite{Abbott:2016blz,*2017-Abbott.etal-APk}.  

The historic detections~\cite{Abbott:2016blz},~\cite{Abbott:2016nmj} were also the first time general relativity was directly tested in strong-field regimes~\cite{TheLIGOScientific:2016src}. This naturally raises the question: 
\emph{Can we (and How to) use QNMs to distinguish between general relativity and modified gravity theories?} For earlier works see Refs.~\cite{PhysRevD.73.064030,*Dreyer:2003bv,*Myung:2011ih,*Myung:2011we}. It is important to note that QNMs bring information about the outer structure of BH spacetimes. While the introduction of the higher derivative terms help to cure the divergence at the curvature singularity, the higher derivative terms can possible change the structure of the horizon. 

In the case of asymptotically flat BHs in general relativity, it is known that two --- odd (Vector) and even (Scalar) --- kinds of gravitational perturbations exist~\cite{1957-Regge.Wheeler-PR,Chandrasekhar1984}. More importantly, the two kinds of perturbations are related to each other and the net emitted gravitational energy is shared equally~\cite{Chandrasekhar1984}.  In this work, we check whether this equality is maintained in the modified theories of gravity. We explicitly show that the QNMs emitted from the BHs in general relativity and $f(R)$ gravity are different and obtain a quantifying tool to distinguish the same.

As mentioned earlier, there is no unique way to modify general relativity~(for recent reviews, see \cite{Clifton:2011jh,Nojiri:2010wj,2010-Sotiriou.Faraoni-RMP,*2010-DeFelice.Tsujikawa-LRR}). In this work, we focus on the simplest possible extension 
of general relativity --- $f(R)$ model:
\begin{equation}\label{fRaction}
S= \frac{1}{2 \kappa^2} \int d^{4}x \, \sqrt{-g} \, f(R) \qquad \kappa^2=\frac{8\pi G}{c^4}
\end{equation}
There are two reasons for this choice: First, they are general enough that higher order Ricci scalar terms can encapsulate high energy modifications to general relativity. Yet the equations of motion are simple enough that it is possible to solve them~\cite{2010-Sotiriou.Faraoni-RMP}. Second and most importantly, $f(R)$ theories do not suffer from Ostr\"ogradsky instability~\cite{2007-Woodard-Proc}. To keep the calculations transparent, we first take $f(R)=R+\alpha R^{2}$, where $\alpha$ is a coupling constant. Later, we extend the analysis to general $f(R)$. 

We use Greek letters for the 4-D space-time, upper case Latin letters for the two angular coordinates and lower case Latin letters for the two orbit coordinates. We follow the notation of Refs.~\cite{2000-Kodama.etal-PRD,*2003-Kodama.Ishibashi-PTP,*2011-Ishibashi.Kodama-PTPS} and set $c=G=1$.
\section{Formalism}
We use the gauge invariant formalism developed in Refs.~\cite{1979-Gerlach.Sengupta-PRD,2000-Kodama.etal-PRD}. For the 4-dimensional 
spherically symmetric space-time, the manifold $ \left(\mathcal{M}, g_{\mu\nu}\right)$ is split into  2-dimensional orbit space $ \left(\mathcal{K}^2, g_{ab}\right) $ and a unit 2-sphere $ \left(\mathcal{S}^2,\gamma_{AB}\right) $. The line element is:
\begin{eqnarray}\label{bgm}
ds^{2} & =&  -g(r)dt^2+\frac{1}{g(r)}dr^2+ \rho^2(r) \, d\Omega^2 
\end{eqnarray}
where $g(r), \rho(r)$ are arbitrary (continuous, differentiable) functions of the radial coordinate $r$.  The metric perturbations about the above background (\ref{bgm}) is given by \cite{Chandrasekhar1984,1979-Gerlach.Sengupta-PRD,2000-Kodama.etal-PRD}
{\small
\begin{equation}\label{ptm}
h_{\mu\nu}dy^{\mu}dy^{\nu}=h_{ab}dx^{a}dx^{b}+2h_{aB}dx^{a}dz^{B}+h_{AB}dz^{A}dz^{B}
\end{equation}
}
Note that under rotations in the 2-sphere, $h_{ab}$, $h_{aB}$, and $h_{AB}$ transform as scalars, vectors, and tensors, respectively. Hence, one can use  
scalar, vector, and tensor spherical harmonics to separate the angular dependence 
of any field appearing in the background or the perturbed space-time~\cite{Chandrasekhar1984}. Imposing the transverse-traceless condition on the tensor perturbations, it can be be shown that the tensor harmonic functions are identically zero~\cite{2013-Cai.Cao-PRD}. Thus, the metric perturbations can be split as scalar and vector perturbations, i. e., $h_{\mu\nu}=h^{S}_{\mu\nu}+h^{V}_{\mu\nu}$. In the linear limit, field equations corresponding to $ h^{S}_{\mu\nu} $ and $ h^{V}_{\mu\nu} $ are decoupled and hence the study of the dynamics of these two type of perturbations can be handled separately~\cite{Chandrasekhar1984}.  For completeness, we have summarized in Appendix \ref{A} and \ref{B}. 

From a combination of metric perturbation variables, two gauge invariant master variables $ \Phi^0_S $ and $ \Phi^0_V $ {for each multipole $ \ell\geq2 $} can be defined that 
satisfy~\cite{2000-Kodama.etal-PRD}  
\begin{eqnarray}
\frac{d^2\Phi^0_S}{dr_*^2}+\left(\omega^2-V_S\right)\Phi^0_S&=&0 \label{mastereqEFV}\\
\frac{d^2\Phi^0_V}{dr_*^2}+\left(\omega^2-V_V\right)\Phi^0_V & =& 0 \label{mastereqEFV2},
\end{eqnarray}
where $r_{*}$ is commonly referred to as tortoise coordinate, $V_S$ and $V_V$ 
are commonly referred to as Regge-Wheeler and Zerilli potentials, respectively~\cite{1957-Regge.Wheeler-PR,Chandrasekhar1984}, and $ \omega $ is the QNM frequency common to both modes. Following points are worth noting: First, the two master variables are not independent, they are related by~\cite{Chandrasekhar1984} 
\begin{equation}
 \Phi^0_{S/V}=\frac{1}{\beta-\omega^2}\left(\mp W\Phi^0_{V/S}+\frac{d\Phi^0_{V/S}}{dr_*}\right) \, , \label{isospectral}
\end{equation}
where $\beta$ is a function of $\ell$ and $W(r)$ is a function of $g(r)$ (See Appendix \ref{C}). Second, the two potentials are also not independent, they are related by~\cite{Chandrasekhar1984,Berti:2009kk} 
\begin{eqnarray}
V_{S/V}=W(r)^2\mp\frac{dW}{dr_*}+\beta \label{isospectralV}
\end{eqnarray}
The above relations hold for vacuum spacetimes and can be extended for other matter sources~\cite{Berti:2009kk}. Third and most importantly, it can be shown that the transmission and reflection coefficients of $V_S$ and $V_V$ are equal for vacuum spacetimes~\cite{Chandrasekhar1984}. Thus, the gravitational radiation from perturbed BHs, as detected at asymptotic spatial infinity, have equal contribution from the scalar and vector modes. Lastly, the equal contribution from the scalar and vector modes may be valid only for general relativity, it may not be valid for modified gravity models. In the rest of this article, we evaluate the contribution of these two types for $f(R)$ gravity model and show that the two contributions are not identical. Thus, QNMs provide a way to distinguish between the general relativity and modified gravity models.
\section{Scalar and Vector modes in $f(R)$}
The equation of motion of (\ref{fRaction}) is
\begin{equation}\label{fieldeq}
G_{\mu\nu}=\kappa^2T^{eff}_{\mu\nu},
\end{equation}
where $ G_{\mu\nu} $ is the Einstein tensor, and
{\small
\begin{equation}
\label{eq:Teff}
T^{eff}_{\mu\nu}=\frac{\alpha}{\kappa^{2}}\left[2\nabla_{\mu}\nabla_{\nu}R-2g_{\mu\nu}\Box R+\frac{1}{2}g_{\mu\nu}R^{2}-2RR_{\mu\nu}\right]
\end{equation}
}
Here, the non linear and higher derivatives of the Ricci curvature are bundled together into the effective Energy Momentum tensor. It is important to note that 
Schwarzschild BH is still the vacuum solution for $f(R)$ model~\cite{delaCruzDombriz:2009et,*Moon:2011hq}.

Perturbing the above field equation (\ref{fieldeq}) about the vacuum solution, we get,
\begin{eqnarray}
& & \delta G_{\mu\nu} = \delta T_{\mu\nu}^{\rm eff} \equiv {2\alpha} \left(\nabla_{\mu}\nabla_{\nu}\delta R-g_{\mu\nu}\Box\delta R\right) \label{ptbfield}\\
&&\Box\delta R-\frac{1}{6\alpha}\delta R=0 \label{traceptb},
\end{eqnarray}
where (\ref{traceptb}) is the trace of (\ref{ptbfield}). Before proceeding to the technical aspects, we would like to mention the following: First, unlike general 
relativity, {gravitational radiation in $f(R)$ gravity has an extra massive scalar degree of freedom which is a mixture of longitudinal and breathing degrees of freedom~\cite{Capozziello:2008rq,Berry:2011pb,Nishizawa:2009bf}. The presence of an extra degree of freedom} can be understood 
by performing a conformal transformation to the metric in action (\ref{fRaction}). Under 
this transformation, $f(R)$ transforms to Einstein gravity plus a scalar field with a potential \cite{MUKHANOV1992203}.  The dynamics of this scalar mode can be found from the trace equation (\ref{traceptb}) as a master equation akin to Eqs.~(\ref{mastereqEFV}) and (\ref{mastereqEFV2}). More specifically, by separating angular dependence from the perturbed Ricci scalar $\delta R=\Omega(x^{a})\textbf{S} $ and making the substitution $\Omega\rightarrow\Phi=r\Omega$ \cite{Nzioki:2014oaa}, we obtain a modified 
Regge-Wheeler equation for a massive scalar field in a Schwarzschild background, i. e.,
\begin{equation}\label{scalmod}
\frac{d^2\Phi}{dr_*^2}+\left(\sigma^2-\tilde{V}_{RW}\right)\Phi=0,
\end{equation} 
where $ \tilde{V}_{RW}=V_{RW}+\frac{g(r)}{6\alpha} $, $ V_{RW} $ is the Regge-Wheeler potential with mass parameter $(6\alpha)^{-1}$, and $ \Phi\left(r,t\right)\equiv \Phi\left(r\right)e^{i\sigma t} $. Thus, in $f(R)$, the extra mode contributes to the emitted gravitational radiation. However, if the extra mode is \emph{not coupled} to the scalar/vector perturbations, 
then it is possible that the scalar and vector perturbations emitted will have 
an equal share, with an small percentage emitted in the extra mode. We 
show this is not the case and that the extra mode couples \emph{only} to
the scalar perturbations. Thus, providing an \emph{unique way} to distinguish $f(R)$ model from general relativity. 

To calculate the modification in the dynamics of scalar and vector perturbations, we need to obtain the corresponding gauge invariant variables  \cite{2000-Kodama.etal-PRD}. From (\ref{eq:Teff}), it is easy to note that 
$T^{eff}_{\mu\nu} $ vanishes for vacuum space-times. Thus, using Stewart-Walker lemma \cite{Stewart49}, the first order perturbed quantity $ \delta T^{eff}_{\mu\nu} $ is gauge invariant. The most general perturbed stress-tensor can be written as~\cite{2000-Kodama.etal-PRD}: 
\begin{widetext}
	\begin{equation}
	\delta T_{\mu\nu}=\left(\begin{array}{c|c}
	\tau_{ab}\textbf{S} & r\tau^{(S)}_{a}\textbf{S}_{B}\\
	---&---------       \\
	r\tau^{(S)}_{a}\textbf{S}_{B} & r^2 \delta P\gamma_{AB}\textbf{S}+ r^2 \tau^{(S)}_{T}\textbf{S}_{AB}
	\end{array}\right)+\left(\begin{array}{c|c}
	0 & r\tau^{(V)}_{a}\textbf{V}_{B}\\
	---&---       \\
	r\tau^{(V)}_{a}\textbf{S}_{B} & r^2 \, \tau^{(V)}_{T}\textbf{V}_{AB}
	\end{array}\right) \label{emptb}
	\end{equation}
\end{widetext}	
Comparing the above expression with $\delta T_{\mu\nu}^{\rm eff}$ in (\ref{ptbfield}), non-vanishing quantities like radial/angular pressure $ \left(\tau_{ab}/ \delta P\right) $, anisotropic stress $ \left(\tau_T\right) $ can be expressed as: 
\begin{eqnarray}
\tau_{ab}=\frac{2\alpha}{\kappa^2}\left[D_{a}D_{b}-g_{ab}\left(\tilde{\Box}+\frac{2}{r}D^{c}rD_{c}-\frac{k^{2}}{r^{2}}\right)\right]\left(\frac{\Phi}{r}\right)& & ~~~\label{emptb1}\\
\tau^{(S)}_{a}= -\frac{2\alpha k}{\kappa^2}D_{a}\left(\frac{\Phi}{r^2}\right); 
\, \,  \tau^{(S)}_{T}=\frac{2\alpha k^{2}}{\kappa^2}\frac{\Phi}{r^3} & & ~~~~~\\
\label{eq:effpressure}
\delta P = \frac{2\alpha}{\kappa^2}\left(\frac{k^{2}}{2r^2}-\tilde{\Box}-\frac{2}{r}D^{a}rD_{a}\right)\left(\frac{\Phi}{r}\right); \, \, k^2 = \ell (\ell + 1) \, , & & ~~~
\end{eqnarray}
while all other quantities vanish. This is the first key result of this work, regarding which we would like to stress a few points:  First,  $\tau_a^{(V)}$ and $\tau_T^{(V)}$ vanish. For $ h^V_{\mu\nu}$, $\delta R$  vanishes, thus, implying that the vector perturbations are not modified. Second, since the radial pressure and anisotropic pressure depend on the extra mode $\Phi$, $\delta R$ for the scalar perturbations do not vanish and the dynamics get modified due to the 
inhomogeneous source term. [For details, see Appendix \ref{D}] Thus, the scalar and 
vector master equations for the spherically symmetric vacuum solution in 
$R+\alpha R^2$ gravity are:
\begin{eqnarray}
&&\frac{d^2\Phi_S}{dr_*^2}+\left(\omega^2-V_S\right)\Phi_S=S^{eff}_S \label{modptbs}\\
&&\frac{d^2\Phi_V}{dr_*^2}+\left(\omega^2-V_V\right)\Phi_V=0 \label{modptbv}.
\end{eqnarray}
Defining $ H(r)\equiv H=k^2+\frac{6M}{r}-2 $ and $ g(r)\equiv g=1-\frac{2M}{r} $ we have,
\begin{eqnarray}
\label{deviation} 
& & \Phi_V = \Phi^0_V \, ; \quad \Phi_S=\Phi^0_S+\tilde{\Phi}_S  \, ; \quad \tilde{\Phi}_S = -\frac{4\alpha}{H}\Phi \, , \label{modDef}\\
& &	S^{eff}_S=  \left[ C_1(\sigma,\omega,r)  
	+ C_2(\sigma,\omega,r)\frac{d}{dr_*} \right] \tilde{\Phi}_S
\end{eqnarray}
details of which are given in Appendix \ref{E}. The coefficients $ C_1 $ and $ C_2 $ are given by
\begin{widetext}
\begin{eqnarray}
& & C_1\left(\sigma,\omega,r\right)=\sigma^2\left(1+\frac{\sigma}{\omega}\right)-\frac{Mg}{r^3}\left(1+\frac{18M}{rH}\right)-\left(\frac{\sigma}{\omega}\right)\frac{g}{r^2}\left[\frac{54M^2}{r^2H}-\frac{72gM^2}{r^2H^2}-\frac{18M}{rH}+\frac{1}{2}\frac{P_1}{H}-\frac{3M}{r}+\frac{\tilde{V}_{RW}}{g}\right]\\
&& C_2\left(\sigma,\omega,r\right)=\frac{3M}{r^2}-\left(\frac{\sigma}{\omega}\right)\left[\frac {12Mg}{r^2H}-\frac {M}{{r}^{2}}\right]; \quad P_1= - \frac{48M^2}{r^2} +  \frac{8M}{r} \left(8 - k^2\right) - 2 k^2 (k^2 - 2)
\end{eqnarray}
\end{widetext}
%
%
Third, Eqs. (\ref{modptbs}) and (\ref{modptbv}) along-with (\ref{scalmod}) form the complete set of equations describing the three propagating degrees of freedom. 
Lastly, due to the change of $\Phi_S$ in (\ref{deviation}), the isospectral relation (\ref{isospectral}) is no more valid (although (\ref{isospectralV}) still holds). Physically this can be understood as the scalar mode leaks energy to the extra mode $\Phi$ via the source term (\ref{modptbs}). Hence, there is an imbalance in the contribution between the scalar and vector types to the emitted gravitational radiation.  

The above analysis can be extended for general $f(R) = \sum_{n = 1}^{\infty} a_n R^n$ model where the effective stress tensor is \cite{2010-DeFelice.Tsujikawa-LRR}:
{\small
\begin{equation}\label{geneffem}
T^{eff}_{\mu\nu}=\frac{1}{\kappa^2 f'}\left[\nabla_\mu\nabla_\nu f'+\frac{g_{\mu\nu}}{2}\left(f-Rf'\right)-g_{\mu\nu}\Box f'\right],
\end{equation}
}
and $ f'(R)=df(R)/dR$. About $ R=0 $, the perturbed stress-tensor becomes:
\begin{equation}
\delta T^{eff}_{\mu\nu}=\frac{1}{\kappa^{2}}\left(\nabla_{\mu}\nabla_{\nu}\delta f'-g_{\mu\nu}\Box\delta f'\right) \label{geneffemptb}
\end{equation}
where $\delta f' = 2\alpha\delta R$.  Substituting $\delta f'$ in (\ref{geneffemptb}) leads to the perturbed stress identical to (\ref{ptbfield}). Thus, all 
the results obtained for $R + \alpha R^2$ gravity is valid for general $f(R)$.

\section{Distinguishing $f(R)$ from GR}

\begin{figure}
\centering
\includegraphics[width=1\linewidth, height=0.24\textheight]{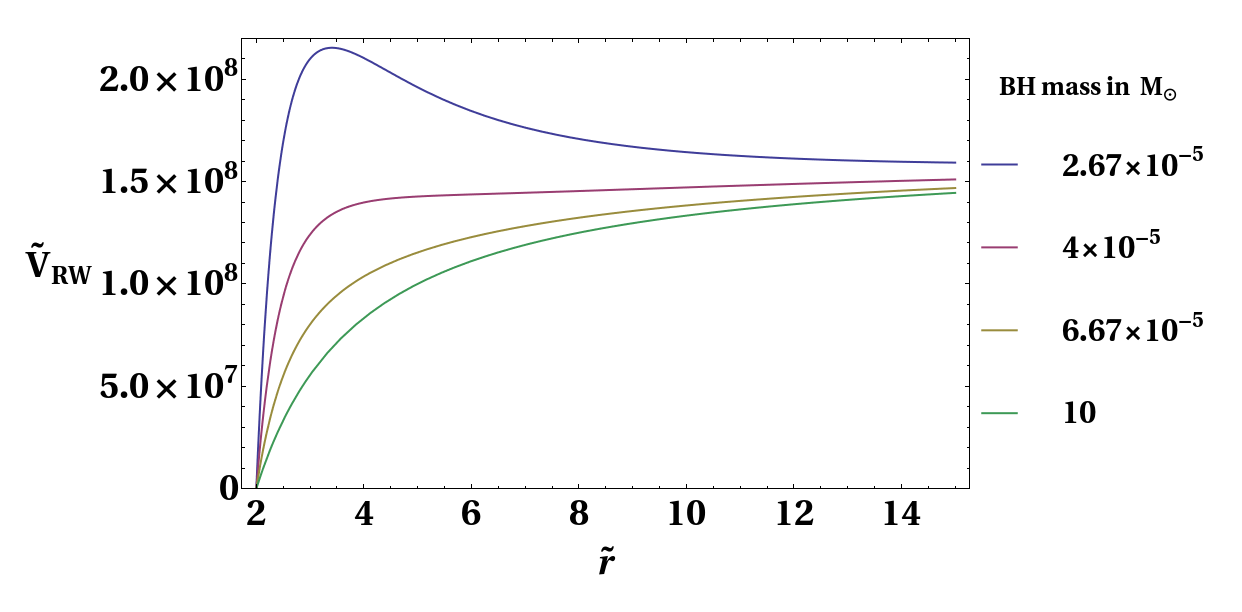}
\caption[Massive scalar mode potential]{Plot of Eq. (\ref{masspot}) for a range of BH masses. For small $ M $ $ \left(\sim10^{-5}M_\odot\right) $, the local maxima reappears, making it similar to the scattering potentials $ V_{S/V} $. For such small masses (which is only possible for primordial BHs), the massive scalar radiation can propagate to $ \infty $ for $ \sigma^2\geq(6\alpha)^{-1} $, and can have a larger share of the net emitted gravitational radiation. For comparison, the green curve shows the potential encountered by $ \Phi $ for a $ 10M_\odot $ BH.}
\label{fig:low_M}
\end{figure}

To go about distinguishing general relativity and $f(R)$ gravity, we need to know $S_{S}^{\rm eff}$ for which we have to obtain the form of the extra mode $\Phi$. To obtain $\Phi$, we need to understand the form of the modified Regge-Wheeler potential: 
{\small
\begin{equation}
\tilde{V}_{RW}\left(\tilde{r}\right)=\left(1-\frac{2}{\tilde{r}}\right)\left[\left(\frac{\ell\left(\ell+1\right)}{\tilde{r}^2}+\frac{2}{\tilde{r}^3}\right)\frac{1}{M^2}+\frac{1}{6\alpha}\right] \label{masspot}
\end{equation}
}
where $ \tilde{r}=\frac{r}{M} $ is a dimensionless radial coordinate. E\"{o}t-Wash experiments provide bound on $\alpha$: $ \alpha \leq 10^{-9} \, m^2 $ \cite{Berry:2011pb,2007-Kapner.etal-PRL}. This is the default 
value used in this study. This leads to a large value of $ \tilde{V}_{RW} $ at infinity, leading to a large potential barrier for the extra mode. Hence, $ \Phi $ does not propagate to $ \infty $, but falls back into the BH after excitation (unless $ \sigma^2\geq(6\alpha)^{-1} $, which becomes the cutoff frequency for the extra mode excitation \cite{Berry:2011pb},\cite{Nzioki:2014oaa}), i.e. a purely decaying solution both in $ r $ and $ t $. For astrophysical-sized BHs and $\alpha \sim 10^{-9} m^2$, Eq.~(\ref{masspot}) can be approximated as
\begin{equation}
\tilde{V}_{RW}\approx\left(1-\frac{2}{\tilde{r}}\right)\frac{1}{6\alpha} \, . \label{masspotapprox}
\end{equation}
which indicates that for observed BH candidates the potential is almost independent of the mass. For small BHs ($ \sim10^{-5} M_{\odot} $), the local maxima reappears (as seen in \ref{fig:low_M}), and the problem becomes a scattering one. We will discuss in the next section, the potential significance of this result for the no-hair theorem~\cite{1995-Bekenstein-PRD,*1998-Bekenstein-Arx}.

An estimate of the maximum deviation from GR can be made by considering the assumption that the intensity of a field (massless or massive) at a point depends inversely on the strength of the potential at that point. Given this, we can define a  ratio of the intensities of the extra mode $ \left(I_{MS}\right) $ and the scalar perturbation $ \left(I_S\right) $ of GR, at a point outside the horizon as:
\begin{equation}
\frac{I_{MS}}{I_{S}}=\frac{V_S}{\tilde{V}_{RW}}=\frac{6\alpha}{M^2}\left[\frac{\ell\left(\ell+1\right)}{\tilde{r}^2}-\frac{6}{\tilde{r}^3}\right], \label{potratio}
\end{equation}
\begin{figure}
	\centering
	\includegraphics[width=\linewidth, height=0.23\textheight]{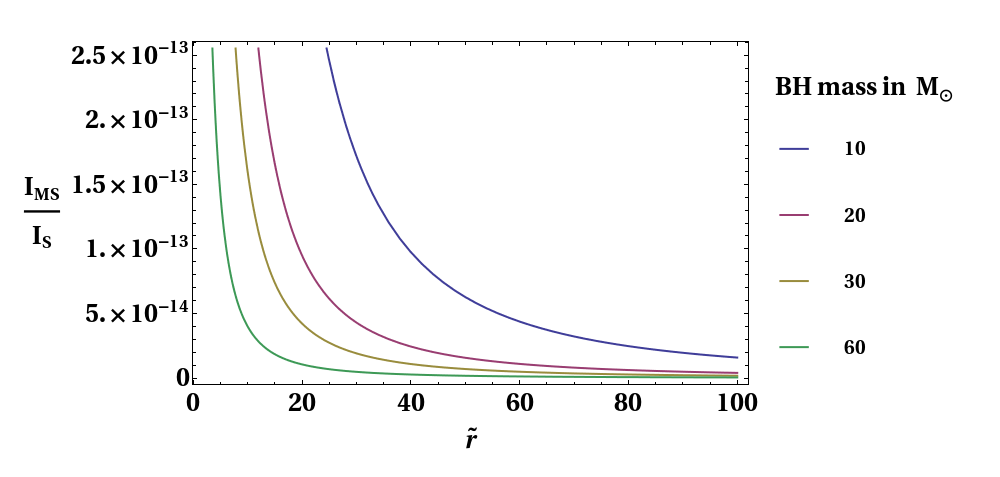}
	\caption{The ratio of the potentials is plotted against the dimensional radius $\tilde{r}$ for different values of M and $ \ell=2 $. The relative intensity is maximum close to the horizon since the extra mode is localized there, dropping away quickly.}
	\label{fig:Potential_ratio}
\end{figure}
where, $\tilde{r}=\frac{r}{M} $. \ref{fig:Potential_ratio} gives a numerical estimate of the relative intensity of the extra mode to the scalar perturbation, at a distance $ \tilde{r}M $ from the horizon. Hence, it is also an upper limit of the relative difference in intensities of the scalar and the vector type perturbations (which is zero in GR). It is important to note that $ V_V $ is used instead of $ V_S $ because of the isospectrality relation between $ V_V $ and $ V_S $ (\ref{isospectralV}). For  $10  M_{\odot}$ BH, $ \ell=2 $, and at a distance of $100 \, M $ from the BH, the relative difference is $ \leq 10^{-14}$. While total deviation can be obtained by integrating from the horizon to infinity i. e. $\sim 10^{-10}$.

\section{Conclusions and Discussions}

Motivated by the historic direct detection of the gravitational waves, 
we have investigated the following question: \emph{How to use QNMs to distinguish 
between general relativity and modified gravity theories}? We have explicitly 
shown that the isospectral relations between scalar and vector perturbations, 
while it holds for general relativity, \emph{does not hold} for $f(R)$ gravity. 
We have shown that the vector perturbations remain unchanged while the scalar
perturbation's modified by the appearance of an extra mode as the source term. 

Physically, the modifications to the scalar perturbations appear by treating the 
extra mode as an effective stress tensor of a perturbed fluid that vanishes 
in the background space-time. The perturbed quantities  (\ref{emptb1})-(\ref{eq:effpressure}) in turn make up an effective source term which couple to the scalar type master equation and modifies the dynamics. The nature of the effective potential (\ref{masspot}) makes the dynamics of the extra mode different from the scalar and vector perturbations. In other words, while the scalar and vector perturbations escape to $ \infty $ as gravitational radiation, the extra mode does 
not. 

The net emitted gravitational energy in $f(R)$ gravity gets shared among the three (scalar, vector and the extra mode) perturbations instead of two perturbations in general relativity. While the vector type retain its contribution, $\Phi $ takes away some the energy released through the scalar type. This in turn also decreases the net gravitational energy that propagates to infinity, owing to the massive nature of the extra mode. Relative intensity of $ \Phi_S $ and $ \Phi $ at any point in the exterior space-time was found to be capped above by  $ \alpha $, indicating the order of the maximum fractional change in energy. At a large, fixed distance $ \left(100M\right) $ from a 10 solar mass BH, we found that the intensity of $\Phi $ relative to $ \Phi_S $ is of the order of $ 10^{-14}$.

The analysis naturally leads to the following question: How this analysis can be used in the current and future gravitational wave detectors to distinguish between 
general relativity and $f(R)$? The change in the intensities of $\Phi_S $ and $\Phi_V $ can be found out from the observations, since a combination of the two manifest as $ + $ and $ \times $ polarization at asymptotic $ \infty $~\cite{Martel:2005ir,Nagar:2005ea}:
\begin{small}
	\begin{eqnarray}
	&& h_+=\frac{1}{r}\sum_{l}\left[\Phi^{(\ell)}_S\textbf{S}_{\theta\theta}^{(\ell)}+\Phi^{(\ell)}_V\textbf{V}_{\theta\theta}^{(\ell)}\right] \label{plus}\\
	&& h_\times=\frac{1}{r\sin\theta}\sum_{\ell}\left[\Phi^{(\ell)}_S\textbf{S}_{\theta\phi}^{(\ell)}+\Phi^{(\ell)}_V\textbf{V}^{(\ell)}_{\theta\phi}\right] \label{cross},
	\end{eqnarray}
\end{small}
details of which were included in Appendix \ref{F}. Using the orthogonality relations of the scalar and vector harmonics~\cite{RevModPhys.52.299}, the above expressions can be rewritten as
\begin{eqnarray}
&& \frac{\Phi^{(\ell)}_S}{r}=\int d\Omega h_+\textbf{S}_{\theta\theta}^{(\ell)}+\int d\Omega \sin\theta h_\times\textbf{S}_{\theta\phi}^{(\ell)} \label{scalar}\\
&& \frac{\Phi^{(\ell)}_V}{r}=\int d\Omega h_+\textbf{V}_{\theta\theta}^{(\ell)}+\int d\Omega \sin\theta h_\times\textbf{V}_{\theta\phi}^{(\ell)} \label{vector},
\end{eqnarray}
where the integration is carried over the two sphere which projects out $ \Phi_{S/V} $ (for each multipole $ \ell $) from the detected polarization. A network of laser interferometers and more detections will be able to measure the individual intensity and angular dependence respectively, of these two polarizations, and hence any difference in the intensities of (\ref{scalar}) and (\ref{vector}) (and hence, the difference in the radiated energy between the two modes) can be calculated.

In Ref. \cite{2013-Arun.Pai-IJMPD}, three model independent tests have been proposed, we are currently investigating a way to use these tests for $f(R)$ model. Our present analysis is in line with the results of~\cite{Konoplya:2016pmh} where it was shown that modifications to gravity could hide in the statistical indeterminacies of parameter matching. In other words, a large class of non-Einstein black holes can mimic the quasinormal spectra of black holes in general relativity. Our analysis is independent of the quasinormal spectra, and the energy shares of the two polarizations of gravitational waves can distinguish between general relativity and $ f(R) $ gravity. We are also currently extending the analysis for rotating and charged black-holes in $f(R)$. 

The issue of broken balance between scalar and vector modes is not unique to $ f(R) $ theories, and hence other modified theories of gravity will show the effect studied in this paper. Moreover, presence of other sources like matter and Electromagnetic fields \cite{Chandrasekhar1984,Gunter497} will also contribute to breaking the balance between the two modes of radiation --- the 
imbalance depending on the distribution of such matter/fields. In general relativity, the energetic difference can be treated as a gravitational measure of \emph{matter+fields} between the source and the detectors. This can be understood by noting that the inhomogeneous source terms appear differentially for general relativity and for f(R) model [See Eqs. (\ref{modptbs}-\ref{modptbv}) and in  RHS of Eqs. (\ref{mastereqEFV}-\ref{mastereqEFV2})]. However, sky observations give us an electromagnetic measure of the \emph{matter and the fields}. Hence, a comparison between electromagnetic and gravitational measure of matter should, in principle, help us put tighter constraints than solar system tests on the parameter $ \alpha $, and consequently, any deviation from general relativity.

Lastly, it is important to note that both in general relativity and $ f(R) $ gravity, the different perturbations have potentials with no local minima. Specifically, in the case of the extra scalar mode, $ \Phi $ can either fall back into the BH or propagate to infinity. It cannot exist in a stable configuration around the BH: which is an indication of the \textit{no scalar hair} \cite{1995-Bekenstein-PRD} property of BH spacetimes. Thus, the uniqueness theorem of general relativity holds true for $ f(R) $ like Lovelock gravity theories in higher dimensions~\cite{Skakala:2013gva}.

\noindent {\it{\bf Acknowledgments}} The authors thank E.~Berti, L.~Cao, V.~Cardoso, A.~Ishibashi, and E.~Poisson for clarifications via email.
The authors thank K.G. Arun and Archana Pai for comments on an earlier draft.
SB is financially supported by the MHRD fellowship at IISER-TVM.  This work is 
supported by Max Planck-India Partner Group on Gravity and Cosmology.
\appendix
\section{Scalar, vector, and tensor harmonics \label{A}}
Any field appearing in the background or the perturbed spacetime can be classified as a scalar, vector or tensor depending on how they transform under rotations in the spherically symmetric spacetime. Thus, angular dependence of any object is separated out using spherical harmonics of three types, which are further subdivided as follows
\begin{itemize}
	\item Angular dependence of a scalar function is expressed as a sum of scalar spherical harmonics $ \textbf{S}\propto Y_{lm}(\theta,\phi) $.
	\item An arbitrary vector field $ U_A $ is expressed as $ U_A=V_A+D_AS $, where $ S $ is a scalar, corresponding to a $ \textbf{S}_A $ (\textit{scalar type vector harmonic}) and a $ \textbf{V}_A $ (\textit{vector harmonic function}).
	\item An arbitrary tensor field $ X_{AB} $ is expressed as $ X_{AB}=T_{AB}+2D_{(A}V_{B)}+\hat{L}_{AB}S+g_{AB}S $, where $ g_{AB} $ is the metric on a $ 2 $ dimensional subspace. $ T_{AB} $ is transverse-traceless tensor field, $ V_A $ is a transverse vector field, projection operator $ \hat{L}_{AB}\equiv D_AD_B-\frac{1}{2}g_{AB} $ obtains a traceless tensor field from a scalar, and $ g_{AB}S $ is the trace part of $ X_{AB} $. These corresponds to $ \textbf{T}_{AB} $ (\textit{tensor harmonic function}), $ \textbf{V}_{AB} $ (\textit{vector type harmonic tensor}), $ \textbf{S}_{AB} $ (\textit{scalar type harmonic tensor}), and $ \gamma_{AB}\textbf{S} $.
\end{itemize}
Out of the seven possibilities listed above, the scalar, vector, and tensor harmonic functions will be called pure and the rest four, derived. The three pure harmonic functions satisfy the following eigenvalue equations
\begin{eqnarray}\label{harmonicdef}
\begin{aligned}
& \left(\hat{\Box}+k^{2}\right)\textbf{S}=0\\
& \left(\hat{\Box}+k^{2}\right)\textbf{V}_{A}=0\\
& \left(\hat{\Box}+k^{2}\right)\textbf{T}_{AB}=0\\
\end{aligned}
\end{eqnarray}
Where $ \textbf{S} $, $ \textbf{V}_A $, and $ \textbf{T}_{AB} $ are scalar, vector, and tensor harmonic functions respectively, and $ k^2=\ell\left(\ell+1\right) $.\\

A vector quantity that can be defined from a pure harmonic function is the scalar type vector harmonic, defined as
\begin{equation}\label{sctv}
\textbf{S}_{A}=-\frac{1}{k}D_{A}\textbf{S}
\end{equation}
$ \textbf{S}_A $ and $ \textbf{V}_A $ form the basis for angular dependency of components transforming as vectors.\\
Similarly, two tensor quantities can be derived which are the scalar and vector type tensor harmonic functions
\begin{eqnarray}
&& \textbf{S}_{AB}=\frac{1}{k^{2}}D_{A}D_{B}\textbf{S}+\frac{1}{2}\gamma_{AB}\textbf{S}\label{sctt}\\
&& \textbf{V}_{AB}=-\frac{1}{2k}\left(D_{A}\textbf{V}_{B}+D_{B}\textbf{V}_{A}\right)\label{vett}
\end{eqnarray}
(\ref{sctt}) and (\ref{vett}) alongwith $ \textbf{T}_{AB} $ and $ \gamma_{AB}\textbf{S} $ form the basis for angular dependence of any fields transforming as tensors.
\section{Vanishing of tensor perturbations on a unit 2-sphere, and its presence on a 3-sphere \label{B}}
\subsection{Trivial $ h^{(T)}_{\mu\nu} $ on $ S^2 $}
The $ S^2 $ background metric is given by
\begin{equation}
g_{AB}\equiv\left(\begin{array}{c c}
1 & 0\\
0 & \sin^2\theta
\end{array}\right)
\end{equation}
On which a transverse and traceless tensor perturbation has the form
\begin{equation}
h_{AB}\equiv\left(\begin{array}{c c}
-\alpha\left(x^a,\theta,\phi\right) & \beta\left(x^a,\theta,\phi\right)\\
&\\
\beta\left(x^a,\theta,\phi\right) & \alpha\left(x^a,\theta,\phi\right)\sin^2\theta
\end{array}\right)
\end{equation}
and satisfies $ D^Ah_{AB}=0 $, which leads to two equations.
\begin{eqnarray}
&&\partial_\phi\left(\beta\sin\theta\right)-\sin\theta\partial_\theta\left(\alpha\sin^2\theta\right)=0\\
&&\partial_\theta\left(\beta\sin\theta\right)+\sin\theta\partial_\phi\alpha=0
\end{eqnarray}
These two lead to the following conditions
\begin{eqnarray}
&&\hat{\Box}\left(\alpha\sin^2\theta\right)=0\\
&&\hat{\Box}\left(\beta\sin\theta\right)=0
\end{eqnarray}
Hence $ \alpha\sin^2\theta $ and $ \beta\sin\theta $ must be proportional to the $ \ell=0 $ harmonic function, which is a constant. Then $ \alpha $ and $ \beta $ should be of the form
\begin{eqnarray}
&&\alpha=\frac{f(x^a)}{\sin^2\theta}\\
&&\beta=\frac{g(x^a)}{\sin\theta}
\end{eqnarray}
For $ \alpha $ and $ \beta $ to be regular at the poles $ \left(\theta=0\right) $, $ f(r) $ and $ g(r) $ needs to go to zero, implying that tensor perturbations on $ S^2 $ vanishes.
\subsection{Non trivial $ h^{(T)}_{\mu\nu} $ on  $ S^3 $ }
The transverse traceless tensor perturbation on a unit 3-sphere can be written as follows
\begin{widetext}
	\begin{equation}
	h_{AB}\equiv\left( \begin {array}{ccc} -\alpha \left( \psi,\theta,\phi \right) &
	\alpha_{{1}} \left( \psi,\theta,\phi \right) &\alpha_{{2}} \left( \psi
	,\theta,\phi \right) \\ \noalign{\medskip}\alpha_{{1}} \left( \psi,
	\theta,\phi \right)  &\alpha \left( \psi,\theta,\phi \right)   
	\sin^2  \psi -\beta \left( \psi,\theta,\phi
	\right)   \sin^2  \psi    &\beta_{{1}}
	\left( \psi,\theta,\phi \right) \\ \noalign{\medskip}\alpha_{{2}}
	\left( \psi,\theta,\phi \right) &\beta_{{1}} \left( \psi,\theta,\phi
	\right) &\beta \left( \psi,\theta,\phi \right)   \sin^2  
	\psi    \sin^2  \theta  
	\end {array} \right) 
	\end{equation}
\end{widetext}
with a background metric
\begin{eqnarray}
g_{AB}\equiv \left( \begin {array}{ccc} 1&0&0\\ \noalign{\medskip}0&  \sin^2
\psi  &0\\ \noalign{\medskip}0&0&  
\sin^2  \psi   \sin^2  \theta
\end {array} \right) \nonumber\\
&& 
\end{eqnarray}
The perturbed metric has 5 functions and 3 equations, whereas for the 2-sphere case the number of functions and equations were equal.\\
$ D^{A}h_{AB}=0 $ corresponds to the following 3 equations
	\begin{eqnarray}
	&&-\sin\psi\sin^2\theta\partial_\psi\left(\alpha\sin^3\psi\right)+\sin^2\psi\left(\sin^2\theta\partial_\theta\alpha_1\right.\nonumber\\
	&&\left.+\sin\theta\cos\theta\alpha_1\right)+\partial_\phi\left(\alpha_2\sin^2\psi\right)=0\\
	&&\nonumber\\
	&&\sin^2\theta\partial_\psi\left(\alpha_1\sin^2\psi\right)-\sin^2\psi\partial_\theta\left(\beta\sin^2\theta\right)\nonumber\\
	&&+\sin^2\psi\left(\sin^2\theta\partial_\theta\alpha+\sin\theta\cos\theta\alpha\right)=0\\
	&&\nonumber\\
	&&\sin\theta\partial_\psi\left(\alpha_2\sin^2\psi\right)+\partial_\theta\left(\beta_1\sin\theta\right)+\sin\theta\sin^2\psi\partial_\phi\beta=0\nonumber\\
	&&
	\end{eqnarray}
$ \alpha_1 $ and $ \alpha_2 $ does not have a derivative-of-a-product form in all the three equations. All the five functions cannot be written individually in the $ \hat{\Box}f=0 $ form. At most, three such relations are possible. This shows how tensor perturbations are present in 1+3 (cosmology) but not in 2+2 (Black Holes).
\section{Isospectrality}\label{C}
\cite{Chandrasekhar1984} found that the potentials $ V_S $ and $ V_V $ have the same reflection and transmission coefficients: which implies that the net emitted gravitational radiation have equal contribution from scalar and vector perturbations. \cite{Chandrasekhar1984} shows that if the two potentials are related to each other through
\begin{equation}
V_{S/V}=W^2\mp\frac{dW}{dr_*}+\beta,
\end{equation}
then the master variables $ \Phi_S $ and $ \Phi_V $ are related as
\begin{equation}\label{iso}
\Phi_{S/V}=\frac{1}{\beta-\omega^2}\left(\mp W\Phi_{V/S}+\frac{d\Phi_{V/S}}{dr_*}\right)
\end{equation}
$ W(r) $ and $ \beta $ are given by
\begin{eqnarray}
&& W(r)=\frac{6M\left(2M-r\right)}{r^2\left(6M+2\lambda r\right)}-\frac{\lambda\left(\lambda+1\right)}{3M} \\
&& \beta=-\frac{\lambda^2\left(\lambda+1\right)^2}{9M^2},
\end{eqnarray}
where $ \lambda=\frac{\left(\ell-1\right)\left(\ell+2\right)}{2} $.\\
The above relations can be generalized to spacetimes with a cosmological constant $ (\Lambda) $ or Electromagnetic fields (as in the case of Reissner-Nordstrom BH) \cite{Chandrasekhar1984}, although they do not hold in general for arbitrary $ T_{\mu\nu} $.

Using (\ref{iso}) and proper boundary conditions (purely ingoing at horizon, purely outgoing at $ \infty $), if $ \omega $ is a QNM frequency of one type, it is also the QNM frequency of the other type. This implies that the scalar and vector perturbations also share the same spectra.
\section{Relating higher derivative terms with the perturbed Energy Momentum tensor \label{D}}
Christoffel symbols of the background metric written in 2+2 form becomes
\begin{eqnarray}
&& \Gamma^a_{bc}=^2\Gamma^a_{bc}\label{chr1}\\
&& \Gamma^a_{BC}=-rD^ar\gamma_{BC}\\
&& \Gamma^A_{aB}=\frac{D_ar}{r}\delta^A_B\\
&& \Gamma^A_{BC}=\hat{\Gamma}^A_{BC}\label{chr4}
\end{eqnarray}
Where $ ^2\Gamma^a_{bc} $ and $ \hat{\Gamma}^A_{BC} $ are Christoffel symbols on $ \mathcal{K}^2 $ and $ \mathcal{S}^2 $ respectively. Using (\ref{chr1})-(\ref{chr4}), various double covariant derivatives were calculated as follows
\begin{eqnarray}
&& \Box F=\tilde{\Box}F+\frac{1}{r^2}\hat{\Box}F+\frac{2}{r}D^arD_aF \label{DD1}\\
&& \nabla_a\nabla_bF=D_aD_bF \label{DD2}\\
&& \nabla_a\nabla_B F=rD_a\left(\frac{1}{r}D_BF\right) \label{DD3}\\
&& \nabla_A\nabla_B F=D_AD_BF+rD^arD_aF\gamma_{AB} \label{DD4}
\end{eqnarray}
for some scalar function $ F(y^\mu) $. The higher derivative terms of the modified field equation, bundled as an effective Energy Momentum tensor has a perturbation around $ R=0 $ given by
\begin{equation}\label{effEM}
\delta T^{eff}_{\mu\nu}=\frac{2\alpha}{\kappa^2}(\nabla_\mu\nabla_\nu\delta R-g_{\mu\nu}\Box\delta R)
\end{equation}
Since $ \delta R $ is a scalar function, its angular dependence was separated by
\begin{equation}
\delta R=\Omega(x^a)\textbf{S}(z^A)
\end{equation}
Using (\ref{DD1})-(\ref{DD4}) in (\ref{effEM}) and using definitions (\ref{sctv}) and (\ref{sctt}) the following relations were obtained
\begin{widetext}
	\begin{eqnarray}
	&& \delta T^{eff}_{ab}=\frac{2\alpha}{\kappa^2}\left[D_aD_b\Omega-g_{ab}\left(\tilde{\Box}\Omega+\frac{2}{r}D^crD_c\Omega-\frac{k^2}{r^2}\Omega\right)\right]\textbf{S} \\
	&& \delta T^{eff}_{aB}=-\frac{2\alpha}{\kappa^2}krD_a\left(\frac{\Omega}{r}\right)\textbf{S}_B \\
	&& \delta T^{eff}_{AB}=\frac{2\alpha}{\kappa^2}\left[k^2\Omega\textbf{S}_{AB}-r^2\gamma_{AB}\left(\tilde{\Box}\Omega+\frac{2}{r}D^arD_a\Omega-\frac{k^2}{2r^2}\Omega\right)\textbf{S}\right]
	\end{eqnarray}
\end{widetext}
$ \tau_{ab} $, $ \tau^{(S/V)}_a $, $ \delta P $, and $ \tau^{(S/V)}_T $ were found by comparing the above relations with the perturbed EM tensor (\ref{emptb})
\begin{eqnarray}
&&\tau_{ab}=\frac{2\alpha}{\kappa^2}\left[D_{a}D_{b}-g_{ab}\left(\tilde{\Box}+\frac{2}{r}D^{c}rD_{c}-\frac{k^{2}}{r^{2}}\right)\right]\left(\frac{\Phi}{r}\right)\nonumber\\
&&\label{emptbA1}\\
&&\tau^{(S)}_{a}=-\frac{2\alpha k}{\kappa^2}D_{a}\left(\frac{\Phi}{r^2}\right)\label{emptbA2} \\
&&\delta P= \frac{2\alpha}{\kappa^2}\left(\frac{k^{2}}{2r^2}-\tilde{\Box}-\frac{2}{r}D^{a}rD_{a}\right)\left(\frac{\Phi}{r}\right)\\
&&\tau^{(S)}_{T}=\frac{2\alpha k^{2}}{\kappa^2}\frac{\Phi}{r^3}\label{emptbA4}\\	
&&\tau_a^{(V)}=0 \label{vec0}\\
&&\tau_T^{(V)}=0 \label{vec1}
\end{eqnarray}
where $ \Phi=r\Omega $ is the extra scalar mode.
\section{The effective source term}\label{E}
The inhomogeneous source term for a general matter perturbation in a Schwarzschild background was obtained for $ m+n $ spacetimes with electromagnetic presence in \cite{2011-Ishibashi.Kodama-PTPS}. Here, we use a restricted version of the above by putting the background and perturbed electromagnetic sources to zero and $ m=n=2 $. We obtain,
\begin{widetext}
	\begin{equation}\label{Seff}
	S^{eff}_S=\frac{g}{rH}\left[ -HS_{T}-\frac{P_1}{H}\frac{S_{t}}{i\omega}-4g\frac{r\left(S_{t}\right)'}{i\omega}-4rgS_{r}+\frac{P_2}{H}\frac{rS^{r}_{t}}{i\omega}+2r^{2}\frac{\left(S^{r}_{t}\right)'}{i\omega}+2r^{2}S^{r}_{r} \right],
	\end{equation}
\end{widetext}
where the prime denotes radial derivative and
\begin{eqnarray}
&& S^a_b=\kappa^2\tau^a_b; \quad S_a=\frac{r\kappa^2}{k}\tau_a^{(S)}; \quad S_T=\frac{2r^2\kappa^2}{k^2}\tau_T^{(S)},\label{Seffcomp}\\
&& P_1= - \frac{48M^2}{r^2} +  \frac{8M}{r} \left(8 - k^2\right) - 2 k^2 (k^2 - 2)\\
&& P_2=\frac{24M}{r}; \quad H=k^2+\frac{6M}{r}-2
\end{eqnarray}
(\ref{Seffcomp}) was calculated using (\ref{emptbA1}),(\ref{emptbA2}), and (\ref{emptbA4}) and substituted to (\ref{Seff}). Time dependence of $ \Phi $ was separated out using
\begin{equation}
\Phi\left(r,t\right)\equiv\Phi\left(r\right)e^{i\sigma t}
\end{equation}
Double radial derivatives were reduced by using the equation of motion of $ \Phi $
\begin{equation}
\frac{d^2\Phi}{dr_*^2}+\left(\sigma^2-\tilde{V}_{RW}\right)\Phi=0
\end{equation}
from which the effective source term was obtained as
\begin{equation}
S^{eff}_S=  \left[ C_1(\sigma,\omega,r) + C_2(\sigma,\omega,r)\frac{d}{dr_*} \right] \tilde{\Phi}_S
\end{equation}
where $ \tilde{\Phi}_S $ was defined in (\ref{modDef}) and the coefficients were obtained as
\begin{widetext}
	\begin{eqnarray}
	& & C_1\left(\sigma,\omega,r\right)=\sigma^2\left(1+\frac{\sigma}{\omega}\right)-\frac{Mg}{r^3}\left(1+\frac{18M}{rH}\right)-\left(\frac{\sigma}{\omega}\right)\frac{g}{r^2}\left[\frac{54M^2}{r^2H}-\frac{72gM^2}{r^2H^2}-\frac{18M}{rH}+\frac{1}{2}\frac{P_1}{H}-\frac{3M}{r}+\frac{\tilde{V}_{RW}}{g}\right]\\
	&& C_2\left(\sigma,\omega,r\right)=\frac{3M}{r^2}-\left(\frac{\sigma}{\omega}\right)\left[\frac {12Mg}{r^2H}-\frac {M}{{r}^{2}}\right]; \quad P_1= - \frac{48M^2}{r^2} +  \frac{8M}{r} \left(8 - k^2\right) - 2 k^2 (k^2 - 2)
	\end{eqnarray}
\end{widetext}
\section{Radiation at infinity} \label{F}
In order to connect the scalar/vector master variables to the radiation detectable at asymptotic spatial infinity, and hence the polarizations, the asymptotic behavior of the perturbed metric was studied by \cite{Martel:2005ir} and found to be
 \begin{equation}\label{asympptb}
 \lim_{r\rightarrow\infty}h_{\mu\nu}\sim\left(\begin{array}{c|c}
 \BigO{\frac{1}{r}} & \BigO{r^0}\\
 ---&---       \\
 \BigO{r^0} & \BigO{r}
 \end{array}\right)
 \end{equation}
Thus, the leading order contribution to gravitational radiation comes only from the $ h_{AB} $ components of the perturbed metric.
$ h_{AB} $ split into scalar and vector perturbations have the following forms
\begin{eqnarray}
&& h^V_{AB}=2r^2H^V_T\textbf{V}_{AB} \label{vec2}\\
&& h^S_{AB}=2r^2\left(H_L\textbf{S}\gamma_{AB}+H_T^S\textbf{S}_{AB}\right)\label{scal2},
\end{eqnarray} 
where $ \textbf{S}, \textbf{S}_{AB} $, and $ \textbf{V}_{AB} $ were defined in Appendix \ref{A}.

(\ref{asympptb}) in a radiation gauge \cite{Martel:2005ir,Nagar:2005ea} should correspond to the TT metric of Minkowski perturbation, from which it can be inferred that only the transverse components of $ h_{AB} $ contribute at leading order at infinity \cite{Martel:2005ir,Nagar:2005ea}. Thus, only $ H^V_T $ and $ H^S_T $ contribute to gravitational radiation. This can be seen as follows: using the definitions of gauge invariant variables in \cite{2000-Kodama.etal-PRD} and enforcing that all other metric components and sources die off faster than $ H_T^{S/V} $, from the definition of master variables and the perturbed field equations, at large $ r $,
\begin{equation}
\Box H^{S/V}_T=0
\end{equation}
were obtained. In Refs. \cite{Martel:2005ir,Nagar:2005ea}, the authors obtain the relation between the metric perturbation and the master variables as
\begin{eqnarray}
&& H_T^S\sim\frac{\Phi^0_S}{r} \label{tscal}\\
&& H_T^V\sim\frac{\Phi^0_V}{2r}
\end{eqnarray}
From the TT metric of Minkowski perturbation, the polarizations can be identified as
\begin{eqnarray}
&& h_+\equiv\frac{h_{\theta\theta/\phi\phi}}{r^2}\\
&& h_\times\equiv \frac{h_{\theta\phi}}{r^2\sin\theta}\label{tpcross}
\end{eqnarray} 
Applying the TT condition to (\ref{vec2}) and (\ref{scal2}), and using (\ref{tscal})-(\ref{tpcross}) we obtain relations between the scalar/vector master variable and the polarizations at each multipole $ \ell $ as
\begin{small}
	\begin{eqnarray}
	&& h_+^{(\ell)}\sim\frac{1}{r}\left[\Phi^{0(\ell)}_S\textbf{S}_{\theta\theta}^{(\ell)}+\Phi^{0(\ell)}_V\textbf{V}^{(\ell)}_{\theta\theta}\right] \label{plusA}\\
	&& h_\times^{(\ell)}\sim\frac{1}{r\sin\theta}\left[\Phi^{0(\ell)}_S\textbf{S}_{\theta\phi}^{(\ell)}+\Phi^{0(\ell)}_V\textbf{V}^{(\ell)}_{\theta\phi}\right] \label{crossA}.
	\end{eqnarray}
\end{small}
Since the extra scalar mode does not travel to $ \infty $, (\ref{plusA}) and (\ref{crossA}) holds for $ f(R) $ gravity as well, with $ \Phi^0_S $ replaced by the modified $ \Phi_S $, whose intensity at the detector is different from $ \Phi^0_V $. 
%
%

\end{document}